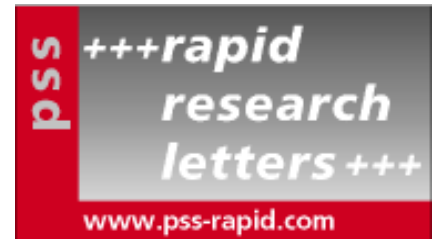

# Doped Nano-Electromechanical Systems

**Dominik V. Scheible[1], Hua Qin[2], Hyun-Seok Kim[2], and Robert H. Blick[*,2]**

[1] Center for Nano-Science, Ludwig-Maximilians-University, Geschwister Scholl Platz 1, 80539 Munich, Germany.
[2] Electrical & Computer Engineering, University of Wisconsin-Madison, 1415 Engineering Drive, Madison, WI 53706, USA.

We present a new generation of nano-electromechanical systems (NEMS), which are realized by doping the semiconductor base material. In contrast to the traditional approach these doped NEMS (D-NEMS) do not require a metallization layer. This makes them far lighter and hence increases resonance frequency and quality factor. Additionally, D-NEMS can be tuned from the conductive state into an insulating one. This will enable a host of new device designs, like mechanically tunable *pin*-junctions and nanomechanical single electron switches. We demonstrate D-NEMS fabrication and operation from the intrinsic, to the light, and to the heavy regime of doping.

**Introduction** The push to realize mechanical devices on the nanoscale has led to the introduction of nano-electromechanical systems (NEMS) [1-4]. Being the next generation emerging after their micron size cousins (MEMS), NEMS promise a broad variety of applications such as ultra-precise sensors and fast mechanical switches. Furthermore, their potential for studying fundamental aspects of quantum electro-mechanics (QEM) is striking [3].

The only major drawback of conventional NEMS is the capacitive actuation, which is bound to metallic layers covering the nanomechanical resonators. Obviously, this delivers impedance matching and sufficient coupling. However, it also increases the mass of the resonators and thus reduces the overall mechanical response, such as the quality factor and the maximal resonance frequency.

Here we demonstrate a way out of this dilemma by using doped-NEMS, or D-NEMS as we termed it. The actual doping is achieved with a state of the art focused ion beam (FIB) writer, which implants Ga-atoms. This enabled us to process standard NEMS resonators *first* – and then dope them locally. Thus we are able to compare the response of undoped and doped-NEMS in their *IV*-characteristics. Furthermore, our work demonstrates the potential FIB-doping has for electromechanical devices. That is a conductive path can be defined within the nanomechanical resonator where needed. This path can then be tuned via the field effect with an appropriate gating electrode. Consequently, *D-NEMS present the next generation of NEMS devices*, which finally will bring the multitude of semiconductor junctions (*pn*, *pin*, *nipi*, and others) to the realm of nanomechanics. To make this point more precise: D-NEMS enables the control of the electric field's potential profile through the alteration of the depletion region by a gating electrode. For conventional NEMS structures, such as resonating beams, this implies that the depletion region can be shifted through the plane of motion of the beam. This constitutes a direct combination of mechanical and electric tunability with unprecedented precision. One can also imagine applications for D-NEMS as mechanical or acoustic logic gates. Such gate structures might then form key elements in mechanical computing circuits [5].

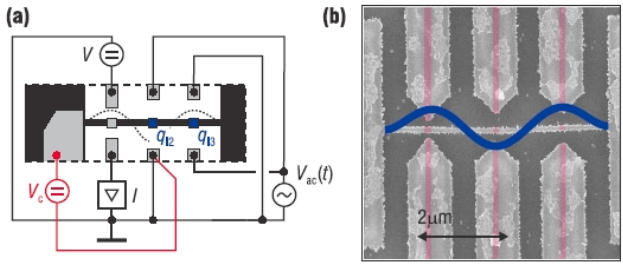

**Figure 1** D-NEMS: underlying red regions indicate the width of Ga-doping. The blue trace indicates the desired mechanical displacement mode.

One of the devices we realized for this experiment is shown in Fig. 1: (a) sketch of the circuit with three islands placed on an undoped beam resonator. The aim is to fabricate an island as a nano-electromechanical single electron transistor (NEMSET), which can exchange charges with the doped leads. While the two islands to the right (blue in Fig. 1(a)) are used to induce mechanical motion via the resonant Coulomb force (RCF) actuation [6] (one of the conventional methods), the third island on the left can exchange electrons with the leads – forming the NEMSET. This is achieved by placing the leads closer to the island than for the other two charge islands. In (b) an SEM micrograph of the actual device is shown. The undoped NEMS resonator is pre-processed from Silicon-On-Insulator (SOI) substrates and locally doped with Ga-ions using a FIB-writer afterwards. Similarly the six gating electrodes are doped, as indicated in Fig. 1(b). The thickness of the single crystalline SOI top silicon layer is 75 nm. The regions of Ga-doping are indicated by the underlying red color. The blue line is a sketch of the desired mechanical displacement mode. We used FIB-deposited Ga-atoms, although Al would be the standard dopant. Ga, however, is preferential here, since Al is not compatible with the NEMS processing wet etch step.

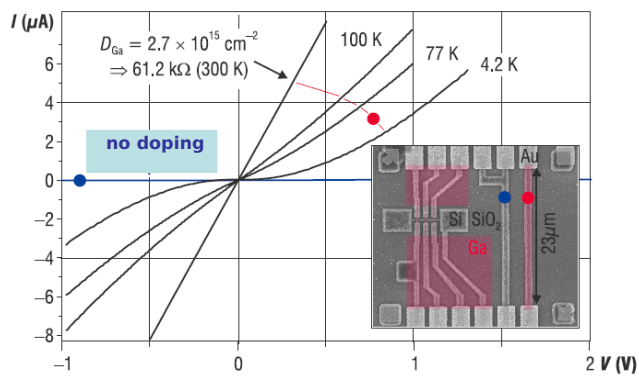

**Figure 2** *IV*-characteristic of two NEMS beam resonators from room temperature to 4.2 K without illumination. The black curves marked by a red dot correspond to the D-NEMS. The undoped-NEMS (blue dot) shows no current flow (blue line). Inset shows the whole processed NEMS structures in top-view with the colors indicating Ga doping (red) and undoped control regions (blue). The contacts are made of Au. The D-NEMS is marked by the red dot and the undoped-NEMS resonator by the blue dot.

We will focus on the first transport measurements on the D-NEMS devices (Fig. 2) by comparing a doped NEMS (red dot) region with an undoped one (blue dot). This is indicated in the inset of Fig. 2: the NEMSET structure from Fig. 1(b) is shown on the left hand side, while we perform transport measurements on the two NEMS resonators on the right hand side.

The nomenclature is as follows: the undoped NEMS we termed M$\pi\pi$M-structure (blue dot), where we assume that the intrinsic Si NEMS resonator can support charge flow under illumination (see below). This M$\pi\pi$M-structure is then comparable to the Schottky solar cell layout (M = metal, $\pi$ intrinsic doping level). On the other hand the local strong doping generates a metal-*p*-metal junction (MpM, *p*-doping by Ga) – our D-NEMS (red dot) with an intrinsic width $W$. The corresponding band diagrams are shown in Fig. 3 for both junction types: (a) gives the MpM- and (b) the M$\pi\pi$M-junction. It should be noted that the D-NEMS doping is very high with a Ga-doping level of $2\times10^{15}$ cm$^{-2}$.

The crucial advantage of D-NEMS is the tunability of the conductive region by an applied voltage. This is commonly expressed by the width $W$ of the Schottky-barrier within the MpM-junction. It is computed from the acceptor concentration $N_A$, the permeability of the semiconductor material $\varepsilon_{Si}\,\varepsilon_0$ and the intrinsic (built in) voltage $V_{bi}$ [7]:

$$W=\sqrt{\frac{2\varepsilon_{Si}\varepsilon_0}{eN_A}(V_{bi}-V)}, \qquad (1)$$

with a dosage used of $N^{2D}_{Ga} = 2\times10^{15}$ cm$^{-2}$ we have in the 100 nm wide Si strip an acceptor density of $N_{Ga} = 2\times10^{26}$ m$^{-3}$ and with Eq.(1) a barrier width of about $W = 1.4$ nm (see Fig. 3). At low temperatures we thus find a symmetric diode-like *IV*-characteristic as shown in Fig. 2, while at room temperature we obtain an ohmic response with a resistance of about 70 k$\Omega$ at room temperature. The undoped or *intrinsic NEMS* (I-NEMS) behaves as an insulator here (blue line), where no illumination is applied.

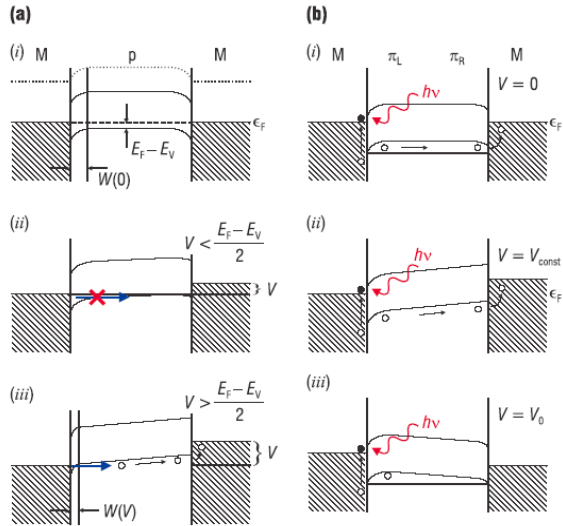

**Figure 3** Band diagrams for the (a) MpM-type and the (b) M$\pi$$\pi$M- structure used here. The wiggled arrows indicate illumination, i.e. induction of charge carriers: (*i*) without voltage applied, (*ii*) below pinch off $V < (E_F-E_V)/2e$ (*ii*) and (*iii*) in the ohmic region. The barrier width of the acceptor layer is determined to be $W = 1.4$ nm.

In order to demonstrate the flexibility of doped NEMS, we now induce charge carriers via illumination, which turned the I-NEMS into a lightly doped D-NEMS. The resulting measurements at 4.2 K are represented in Fig. 4. While the heavily doped MpM-junction shows almost no variation under LED illumination, the M$\pi$$\pi$M-junction reveals a strong effect: without an external voltage the I-NEMS structure acts similar to a Schottky-solar cell, see region (*i*) in Fig. 4. The effect of illumination is best followed in region (*ii*), where we run currents from 1 to 5 mA in steps of 1 mA through the LED. In forward bias, however, in region (*iii*) the now formed light D-NEMS is pinched off. In comparison to the heavy D-NEMS (blue trace) the LED's photons do not alter the *IV*-characteristic of the highly-doped structure, as one would expect. The inset shows the larger scale voltage dependence of the light D-NEMS device after long term illumination.

Finally, let us come back to the initial device design of the doped-NEMSET shown in Fig. 1(b). We have seen above that D-NEMS gives us a broad range of tuning the charge carrier density in the semiconducting material. With the local FIB-deposition of ions we created charged islands within the Si NEMS resonator. We then biased two of the islands with an AC voltage at radio frequencies and induced mechanical motion via RCF. However, the current levels found indicate that even doping to $2 \times 10^{26}$ m$^{-3}$ was not sufficient for forming well-conducting leads. In other words a specific resistance of the Ga-doped Si of $\rho_{Si(Ga)} = 40$ m$\Omega$cm still strongly attenuates the AC voltage of the driving signal. Also, this indicates that the actual doped path width is too narrow. A wider path with a lower resistance will be required to deliver an AC signal to the NEMSET. Nevertheless, we are confident that D-NEMS will be useful for *top-down* fabrication of NEMSETs operating in the few electron limit. This doping of NEMS devices is not limited to semiconductors only. The availability of diamond on SOI wafers will enable similar processes and device realizations in ultrahard materials [8].

In summary we have demonstrated fabrication, characterization, and the use of D-NEMS. This technique now enables tuning the depletion region within the NEMS device for increased mechanical performance. It was also demonstrated that the depletion width can be altered via the bias voltage. Such a variable depletion layer will act as a variable capacitor for exciting NEMS resonators and can make pure metallic layers obsolete. This will considerably enhance the mechanical properties of NEMS. As we have shown FIB-deposition of dopants has the additional advantage of creating a conductive path within a NEMS device. In more detail we have presented the different regimes of doping NEMS, starting with intrinsic NEMS, to light and heavy doping towards the metallic limit.

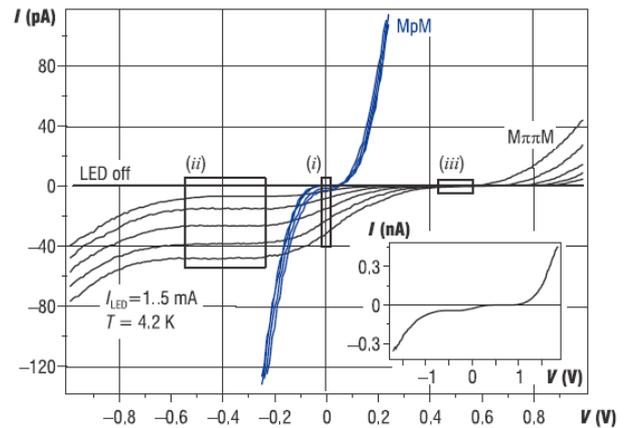

**Figure 4** *IV*-characteristic of the highly-doped NEMS structure (MpM, blue traces) compared to the undoped device (M$\pi$$\pi$M, black traces) under illumination. The measurements are performed at 4.2 K with the LED bias with a current from 0 – 5 mA in 1 mA-steps. The inset shows the intrinsic-NEMS (undoped) voltage dependence over a larger range after illumination. Well pronounced are the exponential onsets of current flow.

**Acknowledgements** We like to thank DARPA for support through the NEMS-CMOS program and the Wisconsin Alumni Research Foundation (WARF) for partial support of this work. We like to thank SOITEC's George Celler for support. We also thank A.D. Wieck, S. Hoch, and A. Ebbers for support with focused ion beam writing.